\documentclass[%
 reprint,
 amsmath,amssymb,
 aps,
prl,
]{revtex4-1}

\usepackage{graphicx}
\usepackage{dcolumn}
\usepackage{bm}
\usepackage{comment}
\usepackage{notoccite}

\begin{document}

\title{Disappearance of the hexatic phase in a binary mixture of hard disks}

\author{John Russo}
\email{john.russo@bristol.ac.uk}
\affiliation{%
School of Mathematics, University of Bristol, Bristol
BS8 1TW, United Kingdom 
}%

\author{Nigel B. Wilding}%
 \email{n.b.wilding@bath.ac.uk}
\affiliation{%
 Department of Physics, University of Bath, Bath BA2 7AY, United Kingdom}

\date{\today}

\begin{abstract}

Recent studies of melting in hard disks have confirmed the existence of a hexatic phase occurring in a narrow
window of density which is separated from the isotropic liquid phase by a first-order transition, and from the solid
phase by a continuous transition. However, little is known concerning the melting scenario in mixtures of hard disks.
Here we employ specialized Monte Carlo simulations to elucidate the phase behavior of a system of large ($l$) and small
($s$) disks with diameter ratio $\sigma_l/\sigma_s=1.4$. We find that as small disks are added to a system of large
ones, the stability window of the hexatic phase shrinks progressively until the line of continuous transitions
terminates at an end point beyond which melting becomes a first-order liquid-solid transition. This
 occurs at surprisingly low concentrations of the small disks, $c\lesssim 1\%$, emphasizing the fragility of the hexatic phase.  We speculate that the change to the melting scenario is a consequence of strong fractionation effects, the nature of which we elucidate.

\end{abstract}

\maketitle

One of the most celebrated accomplishments of statistical mechanics is the progress in understanding the rich physics of
phase transitions in two-dimensional (2D) systems. In particular, the melting of 2D crystals has puzzled researchers for
decades. Early theoretical considerations~\cite{peierls1934bemerkungen} seemed to rule out the existence of 2D solids,
while the celebrated \emph{Mermin-Wagner} theorem~\cite{mermin1966absence,mermin1968crystalline} proved rigorously that
short-range continuous potentials cannot possess long-range positional order~\cite{hohenberg1967existence}. However,
these theoretical ideas were in conflict with early simulation results for hard disks~\cite{alder1962phase}
which suggested the presence of a first-order phase transition between a liquid and a solid \cite{alder1962phase}. The
early simulations established hard disks as a benchmark system for testing theories of 2D melting, and
motivated~\cite{kosterlitz2016kosterlitz} Kosterlitz and Thouless (KT) to develop the theory that now bears their
name~\cite{kosterlitz1972long} (also independently found by Berezinskii~\cite{berezinskii1971destruction}). Within the
KT theory, a new type of 2D solid phase is proposed, having long-range orientational order and only quasi-long range
positional order, and whose melting involves the continuous unbinding of dislocation pairs. The phase that results from the KT
transition mechanism was originally believed to be an isotropic liquid, but Nelson,
Halperin~\cite{halperin1978theory,nelson1979dislocation}, and Young~\cite{young1979melting}, realized that the new phase
retains quasi-long range orientational order, and melts via a second KT transition involving the unbinding of
dislocations into free disclinations. The intermediate phase was called the \emph{hexatic phase}, and the scenario of
melting via two continuous transitions is known as the KTHNY theory.

The KTHNY theory is based on the assumption that the solid phase remains stable on decompression until the continuous dislocation unbinding
transition. Consequently it does not rule out the possibility that this transition is pre-empted by a first-order transition, as simulations at first seemed to suggest. The two competing scenarios were debated for decades~see e.g.
\cite{strandburg1988two,Weber1995,dash1999history,jaster1999computer,mak2006large}, until a new class of rejection-free
algorithms was developed by Bernard and Krauth, called \emph{event-chain} algorithms, that allowed the simulation of large
systems in the melting region. This led to a surprising discovery~\cite{bernard2011two}: the melting of hard disks
occurs via a continuous KT transition between the solid and hexatic phase, and via a first-order transition between the
hexatic and liquid phase. The work has been extended and generalized to soft potentials~\cite{kapfer2015two}, to hard
polygons~\cite{anderson2017shape}, and to hard-sphere monolayers~\cite{qi2014two}, where the findings were even verified
experimentally for colloidal particles~\cite{thorneywork2017two}.

Hitherto, most studies of 2D melting have focused on pure systems. However, many important natural and technological systems
are {\em mixtures} of different sized particles, and exhibit phase behaviour that is far richer and more complex than
for pure systems. A specific question of fundamental interest is: ``what happens to the melting transition of a system of pure hard
disks when a low concentration of smaller hard disks is added?'' This second species acts as a form of
disorder which can selectively favor one particular phase, and might change the nature of the
transition~\cite{kindt2015grand}. In this Letter we consider the melting scenario for such a binary system of hard disks. We choose the size
ratio between large ($l$) and small ($s$) disks to be $\sigma_l/\sigma_s=1.4$, which is large enough to constitute a significant perturbation,  but
small enough to ensure that the minority ($s$) component is included substitutionally rather than interstitially in the solid phase. 
We define the concentration of small disks by $c=N_s/(N_l+N_s)$, with $N_l$ and $N_s$ the number of large and small disks respectively.

\begin{figure*}[!ht]
 \centering
 \includegraphics[width=15cm]{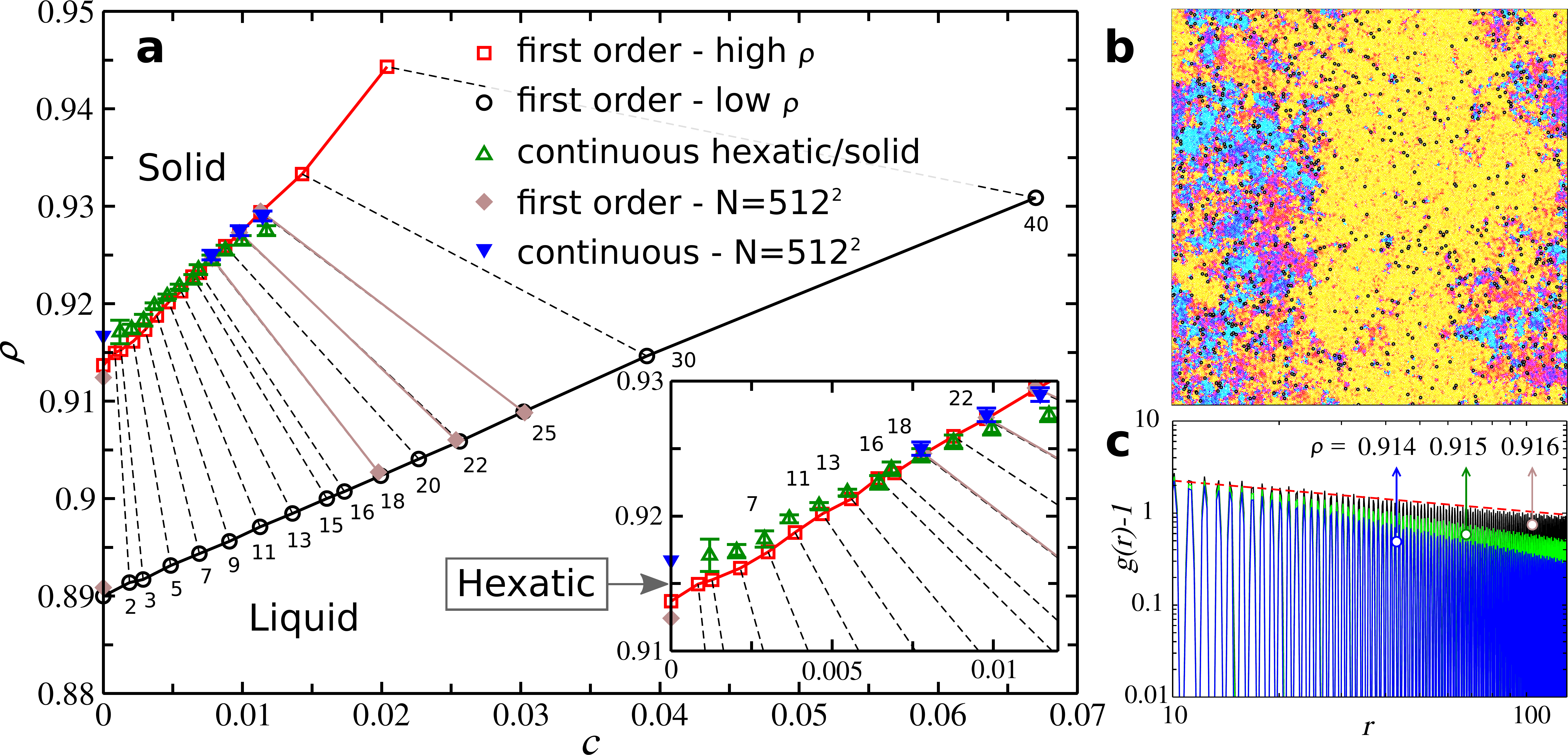}
 
\caption {{\bf (a)} Phase diagram in the $\rho$-$c$ plane as discussed in the text.
 Dashed tie lines connect first-order coexistence points for various (marked) fugacity fractions ranging from  $\xi=2$ to
$\xi=40$. Also shown is the line of continuous hexatic-solid transitions. Error bars were determined as described in the SM \cite{Russo:SM2017}.  Data correspond to $N=256^2$ unless otherwise indicated. {\bf (b)} Snapshot in the
coexistence region at $\xi=15$ and $\rho=0.91$. The color refers to the phase of the hexatic order parameter,
while unfilled circles are small disks plotted for clarity with twice their true size. {\bf (c)} Radial distribution function obtained by projecting  the pair correlation function $g(\Delta\mathbf r)$ onto the direction of a lattice vector. Data is shown for $\xi=2$ at $\rho=0.914$ and $0.915$ (hexatic phase), and $\rho=0.916$ (solid phase). The dashed line is the power-law scaling predicted by KTHNY theory.} 
\label{fig:phase}
\end{figure*}

In order to correctly obtain accurate coexistence properties of mixtures, one needs to carefully account
for fractionation effects ie. the different partitioning of species among the coexisting
phases~\cite{wilding2010phase}. Open ensembles are particularly suited for this purpose as they permit global
concentration fluctuations. Here we employ Monte Carlo (MC) simulation in the semi-grand canonical Monte ensemble
(SGCE), in which a randomly selected disk can change its species \cite{frenkel2001understanding} under the control of a
fugacity fraction $\xi=f_s/(f_l+f_s)$, where $f_s$ and $f_l$ are the fugacities of the $s$ and $l$ disks respectively.
The value of $\xi$ sets the overall concentration which (due to fractionation) will generally differ from that of
the individual phases. In order to accelerate local density and concentration fluctuations we implement the
event-chain algorithm~\cite{bernard2011two,michel2014generalized}, as well as a MC position swap for randomly
selected pairs of disks. In the regime of small $c$ of interest here, $\xi$ is of the order of $10^{-5}$ and for
convenience we quote only its coefficient, e.g. we write $\xi=3$ instead of $\xi=3\times 10^{-5}$.

The hard disks occupy a periodic square box of side $L$ which, in common with other lengths, we measure in units of
$\sigma_l$. In order to study the effects on the phase behavior of varying $c$, we selected $16$ different values of $\xi$ in the interval
$\xi\in[0,40]$. For each $\xi$, we scanned (in a stepwise fashion) the range of number density $\rho=(N_s+N_l)/L^2$ over
which melting occurs. Within a two-phase region these paths of constant $\xi$ correspond to \emph{tie lines} along which phase separation occurs
at constant fugacity for each component. We measured the pressure $P$ along each tie line to obtain the corresponding
equation of state (EOS), $P(\rho)$. This was found to exhibit the typical \emph{van der Waals loop} of a first-order phase transition in a finite-sized system~\cite{binder2012beyond},
which for pure hard disks corresponds to the liquid to hexatic transition~\cite{bernard2011two}.  
Application of the Maxwell construction to the EOS permitted the determination of the coexisting densities that mark the termini of the tie line.

In order to locate the density of the KT transition separating the hexatic and solid phases, we extended to binary
mixtures the methods introduced in Ref.~\cite{bernard2011two}. Specifically, for each $\xi$ we computed the form of the
pair correlation function $g(\Delta\mathbf r)$ for a sequence of values of density. The hexatic-solid transition is
signaled by a crossover in the form of $g(\Delta\mathbf r)$ from exponential (short-range, hexatic) order to power-law
(quasi-long range, solid) behavior. Since 2D systems of disks can exhibit a very large correlation length, the accurate
location of this crossover density requires simulations of considerable size and duration. Most of our estimates of phase
behavior were performed, using $N=N_s+N_l=256^2$, while $N=512^2$ was used for a selected number of fugacities
in order to assess finite-size effects. Overall, our simulations consumed well over $100$ years of
single-core CPU time.

Fig.~\ref{fig:phase}(a) presents our measurements of the density-concentration phase diagram. Apparent is a region of
first-order phase coexistence delineated by coexistence state points connected by dashed tie lines. Within this region a
lower density phase coexists with a higher density phase, the nature of which we now examine for a moderate value of $\xi$. A configurational
snapshot inside the coexistence region at $\xi=15$ and $\rho=0.91$, is displayed in Fig.~\ref{fig:phase}(b) and shows
small disks as unfilled circles, while large disks are colored according to the phase of the hexatic order parameter,
$\psi^j_6=\sum_k\exp(i6\theta_{jk})/n_j$, where, for each disk $j$, $k$ is one of the $n_j$ nearest neighbors (defined
as the disks whose cells share one edge with $j$ in the radical Voronoi tessellation of the plane), and $\theta_{jk}$ is
the angle that the vector $r_{jk}$ makes with a reference direction. Evident is a strip of dense phase having strong
hexatic ordering which is separated by a rough interface from a  disordered (liquid) phase of lower density. The supplementary material (SM)\cite{Russo:SM2017}
additionally shows that the EOS  displays a van der Waals loop. Both these properties are the
hallmarks of a first-order phase transition. Our phase diagram shows that on adding small particles, the region of
first-order coexistence (as found for the pure system in Ref.~\cite{bernard2011two}) extends along the concentration axis,
moving to higher $\rho$ (and higher $P$). Interestingly, as $\xi$ increases, the tie lines lengthen while their slope
rapidly flattens, implying that small disks are more easily `dissolved' in the disordered liquid phase. Our system thus
behaves like a \emph{eutectic mixture}.

Also indicated in Fig.~\ref{fig:phase}(a) is the line of continuous hexatic-solid transitions, marked by the green and blue triangles, determined from the
crossover in the form of the decay of the pair correlation function. The nature of this crossover \cite{Russo:SM2017} is shown in Fig.~\ref{fig:phase}(c)
which plots $g(r)-1$ for state points spanning the transition line. The blue curve exhibits exponential decay (albeit with
a very long correlation length) characteristic of the hexatic phase, while the black curve exhibits power-law decay
characteristic of the 2D solid. The power-law exponent is compatible with $-1/3$ (red dashed line), which
corresponds to the predicted stability limit of the solid phase within KTHNY theory. Our results show that, as small
particles are added to the system, the continuous hexatic-solid transition point of the pure system becomes a line of KT
transitions that extends to higher densities.

The inset of Fig.~\ref{fig:phase}(a) expands the high density region of the phase diagram, revealing that as $\xi$ is
increased, the window of stability of the hexatic phase shrinks. For $\xi=20\pm 2$, corresponding to concentration
$c\approx 1\%$, the KT line {\em intersects} --and extends metastably into-- the region of first-order coexistence.
Thereafter, for $\xi\gtrsim 20$, the liquid phase coexists with a solid rather than a hexatic phase. Such an
intersection point is analogous to the critical end point that features in the phase diagrams of many binary mixtures
\cite{Fisher:1990aa,wilding1997b}. There a line of critical demixing transitions intersects and is truncated by a first-order transition line, the latter inheriting the singularities of the former. However, a difference between a critical end point and the end point in the present system
is that the KT transition is a phase transition of infinite-order within the classification scheme of Ehrenfest, and
thus the free energy and all its derivatives are continuous. Accordingly one expects that the first-order boundary in
our system remains analytic.

In order to check for finite-size effects with regard to the phase behaviour, we performed simulations with $N=512^2$
for $\xi=0,18,22,25$, the results of which are included in Fig.~\ref{fig:phase}(a). As in Ref.~\cite{bernard2011two},
the first-order coexistence window for the pure system is observed to shrink noticeably with increasing system size, but
this narrowing is much reduced at high $\xi$ where the transition points are indistinguishable within our precision. The
corresponding results for the KT line shift slightly to higher $\rho$ with increased system size, but the results are
still within the error bars. Overall, therefore, the $N=512^2$ results confirm the shrinking of the hexatic stability
window.

Further evidence corroborating the disappearance of the hexatic phase with increasing $\xi$ can be gleaned from a study
of the elastic constants of the solid phase. More specifically, we exploit KTHNY predictions for the value of the Young's
modulus ($K$) at the limit of stability of the solid phase, to independently deduce the locus of the KT line. Our approach
follows Refs.~\cite{sengupta2000elastic,sengupta2000elastic2} and involves measurements of the Lagrangian strain tensor in
combination with a finite-size scaling analysis~\cite{sengupta2000elastic} to yield estimates for the bulk and the
effective shear moduli in the thermodynamic limit. Since this approach applies only to defect-free solids and is feasible only for
much smaller system sizes than considered above, we implement it in SGCE MC simulations of $N=3120$ disks in which the
generation of defects is suppressed by the simple expedient of rejecting any MC update that would create a dislocation
pair of the smallest Burger's vector. The resulting elastic constants provide an estimate of $K$ for the constrained
(ie. defect-free) solid, which we measure as a function of $\rho$ for a number of values of $\xi$.

\begin{figure}
 \centering
 \includegraphics[width=8cm]{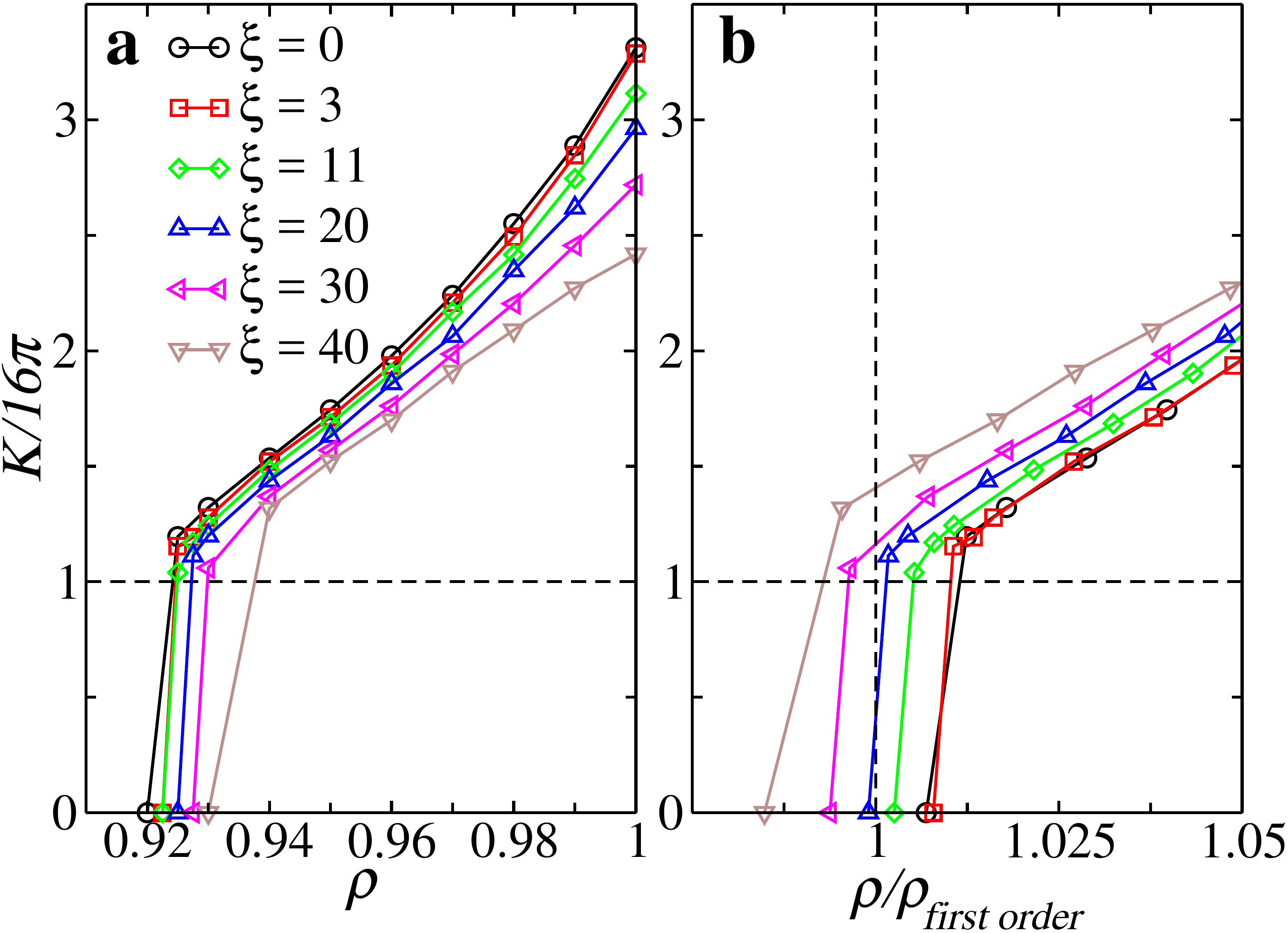}
\caption{{\bf  (a)} Normalized Young modulus ($K/16\pi$) for various $\xi$ as a function of $\rho$. The horizontal lines marks the stability limit  $K=16\pi$ for the solid phase.
{\bf (b)} The same data plotted as a function of $\rho/\rho_\text{first order}$,  the high density boundary of the first-order transition. The point with coordinates $(\rho/\rho_\text{first order}=1,K/16\pi=1)$  marks the end point of the KT line (see text). }
\label{fig:young}
\end{figure}
 
Of course, equilibrium configurations of an unconstrained solid actually contain a non-zero population of defects, and
consequently the measurements of $K$ must be corrected to take this into account. KTHNY theory
\cite{nelson1979dislocation} provides a framework for doing so which involves solving a set of renormalization group
recursion relations~\cite{sengupta2000elastic2}, starting from the $K$ value of the defect-free configurations. The
solution also requires an independent calculation of the fugacity of dislocation pairs $y$. As shown in
Ref.~\cite{sengupta2000elastic2}, this quantity is proportional to the acceptance probability in the constrained
ensemble simulations, which we match with the site-fraction $N_i/N$ of dislocations pairs in the monodisperse system
(see SM~\cite{Russo:SM2017}). The corrected values of $K$ for different values of $\xi$ are plotted in
Fig.~\ref{fig:young}(a). Note that KTHNY theory predicts the melting of the solid to occur when $K/16\pi$ reaches unity
under decompression, whereupon $K$ jumps abruptly to zero. Fig.~\ref{fig:young} thus shows that the effect of adding
small disks is to increase the softness of the solid, i.e. lowering $K$ with increasing $\xi$, and moving the KT
transition density from $\rho=0.92$ for the pure case ($\xi=0$) to $\rho=0.935$ for $\xi=40$.

Overall, the results emerging from Fig.~\ref{fig:young}(a) for the $\xi$ dependence of the KT transition density are in
good agreement with the results deduced from the pair correlation function that are shown in Fig.~\ref{fig:phase}(a).
Additionally, the data for $K(\rho)$ confirms the disappearance of the hexatic phase. This can be appreciated by replotting $K$ as a function of $\rho/\rho_\text{first order}$, with $\rho_\text{first
order}$ the high density boundary of the first-order transition at each value of $\xi$. Represented in this way, the
data (Fig.~\ref{fig:young}(b)) reveal that the shift of the KT transition to higher density with increasing $\xi$ is
{\em slower} than the shift of the first-order boundary, which therefore ultimately engulfs it. The value of $\xi$ for which the 
KT line reaches its endpoint can be read off from Fig.~\ref{fig:young}(b) by locating that curve which intersects the point
whose coordinates are $\rho/\rho_\text{first order}=1, K/16\pi=1$. Notwithstanding a certain
sensitivity to the method by which the fugacity of dislocation pairs is determined, we estimate this to occur for
$\xi\cong 20$, in accord with the previous estimate shown in Fig.~\ref{fig:phase}(a).

\begin{figure}
 \centering
 \includegraphics[width=8cm]{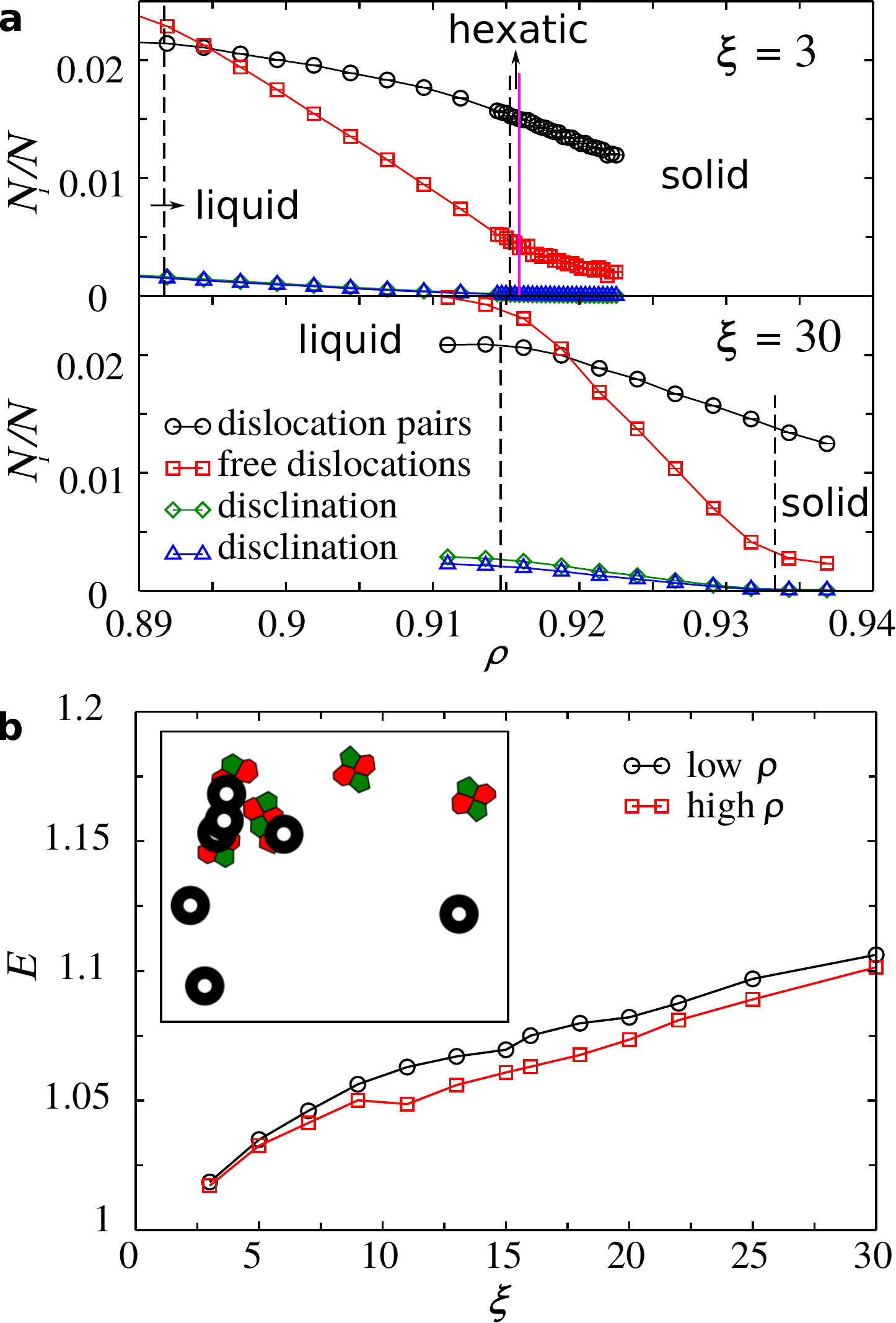}
\caption{{\bf  (a)} Fraction of sites with defects for $\xi=3$ (top panel) and $\xi=30$ (bottom panel). The dashed vertical lines represent the boundaries of the first-order phase transition. The vertical
(purple) continuous line in the top panel marks the location of the continuous transition.
{\bf (b)} $\xi$ dependence of the Pielou evenness index ($E$) for state points at the low and high density boundary of the first-order coexistence region. The inset shows a snapshot for a small portion of a system at $\xi=30$ and $\rho=0.9343$, where we plot both small disks (at twice their real size) and
Voronoi cells representing topological defects with 5 (green cells) and 7 (red cells) neighbors. Large disks are omitted.
}
\label{fig:defects}
\end{figure}

We now attempt to rationalize the loss of the hexatic phase. To
this end we search for changes in the structural character of the phases that might affect their entropy balance. We
focus on three properties: (i) defect populations, (ii) spatial correlations between small particles, and (iii)
degree of fractionation. Fig.~\ref{fig:defects}(a) plots the density dependence of the population of those defects
relevant for the KTHNY transition namely $5-7$ dislocation pairs, free $5-7$ dislocations and disclinations with $5$ and
$7$ neighbors. Results are shown for $\xi=3$ (top panel) and for $\xi=30$ (bottom panel) \footnote{We consider only
dislocation pairs of the smallest Burger's vector, as these are the most abundant at the densities considered.}.
Vertical dashed lines mark the coexistence boundary for the first-order transition, and for $\xi=3$ (top panel) the
purple line marks the density of the continuous transition. One observes that the KTHNY sequence of dislocation pairs
(solid) $\rightarrow$ free dislocations (hexatic) $\rightarrow$ free disclinations (liquid) is obeyed for all $\xi$
considered. One also notes from Fig.~\ref{fig:defects}(a) that on traversing the phase transitions, the site fraction
($N_i/N$) of all topological defects remains practically unchanged for both low and high $\xi$ (a fact that is confirmed
by the snapshots shown in Fig. S7 of the SM~\cite{Russo:SM2017}).

In order to assess how small particles are distributed within the system, we plot in Fig.~\ref{fig:defects}(b) their
\emph{Pielou} index ($E$) as a function of $\xi$ for densities on the low and high density boundary of the first-order
coexistence region. This quantity measures the \emph{evenness} of a distribution of points on a
plane~\cite{pielou1966measurement}: $E=\pi\rho_s\bar\omega$, where $\rho_s$ is the number density of small disks and
$\bar\omega$ is the average squared distance between a randomly chosen point on the plane and the nearest small disk.
$E=1$ signifies a random (Poisson) distribution, while $E>1$ indicates clustering. As Fig.~\ref{fig:defects}(b) shows,
the value of $E$ in both the liquid and the high density phase grows strongly with increasing $\xi$, indicating an
increase in the clustering of small particles. Such clustering is confirmed visually by the snapshot shown in the inset
of Fig.~\ref{fig:defects}(b) which depicts a small area of the solid phase for $\xi=30$ (a larger portion is shown in
Fig.~S7 of the SM~\cite{Russo:SM2017}). Interestingly, however, the $E$-index of the two phases are very similar for a given $\xi$, despite their very different values of $c$ and degree of structural order. This
latter finding suggests that the contribution of the clustering of small particles to the entropy of the solid and
liquid phase are rather similar. We also note in passing that small particle clusters tend to occur together with clusters of defects as can be seen in the inset of Fig.~\ref{fig:defects}(b), as well as Fig S7
of the SM \cite{Russo:SM2017}) 

Taken together, this evidence shows that (i)  the addition of small particles seems to have negligible effects on the population of defects at the phase transitions;
(ii) while the small particles cluster quite strongly at large $\xi$, there is little {\em difference} in the degree
of small particle clustering between the phases (which might otherwise alter their entropy
balance). Accordingly, it is not clear that one can attribute the loss of the hexatic to either changes in
defect populations or to particle clustering. Instead we speculate that the effect is primarily driven simply
by fractionation. Specifically, as $\xi$ increases, the small particles show a strong preference to migrate to the liquid phase. This in turn
raises the (mixing) entropy of the liquid much more than it does that of the hexatic or the solid phase. The result is to stabilize the liquid
relative to the higher density phase and hence to shift the first-order coexistence region to higher density faster
than the KT line.  Ultimately, then, the liquid region engulfs the hexatic, leaving only liquid-solid coexistence in its stead.

To conclude, using tailored large-scale MC simulations, we have investigated the melting scenario of a binary mixture of
hard disks with $\sigma_l/\sigma_s=1.4$ as a function of the fugacity fraction $\xi$ of the small particles. We have
shown that the hexatic phase that occurs in the pure case survives only for very low concentrations of small particles
$c \lesssim 1\%$, demonstrating that it is an utmost delicate state of matter. For larger $c$ the melting scenario
changes into a first-order transition between the liquid and solid phases, a finding which we have argued is a corollary of the
stabilizing effect of strong fractionation of small particles to the liquid phase.

\noindent
{\bf Acknowledgments} We thank H. Tanaka and R. Evans for helpful discussions. JR thanks the Royal Society for a URF fellowship, 
and the Blue Crystal supercomputer at the University of Bristol for a generous allocation of resources. Part of this research made
use of the Balena High Performance Computing Service at the University of Bath.

\bibliography{biblio}
\newpage

\setlength{\tabcolsep}{10pt}

\newenvironment{sistema}%
  {\left\lbrace\begin{array}{@{}l@{}}}%
  {\end{array}\right.}

\renewcommand{\figurename}{Fig.} 
\renewcommand{\thefigure}{S\arabic{figure}} 


\onecolumngrid

\centerline{\bf \large Supplementary Information for}
\centerline{\bf \large ``Disappearance of the hexatic phase in a binary mixture of hard disks''}
\vspace{0.5cm}


\noindent

\section{First-Order Transition: van der Waals loop}
\label{sec:fo}

\begin{figure}[!ht]
 \centering
 \includegraphics[width=10cm]{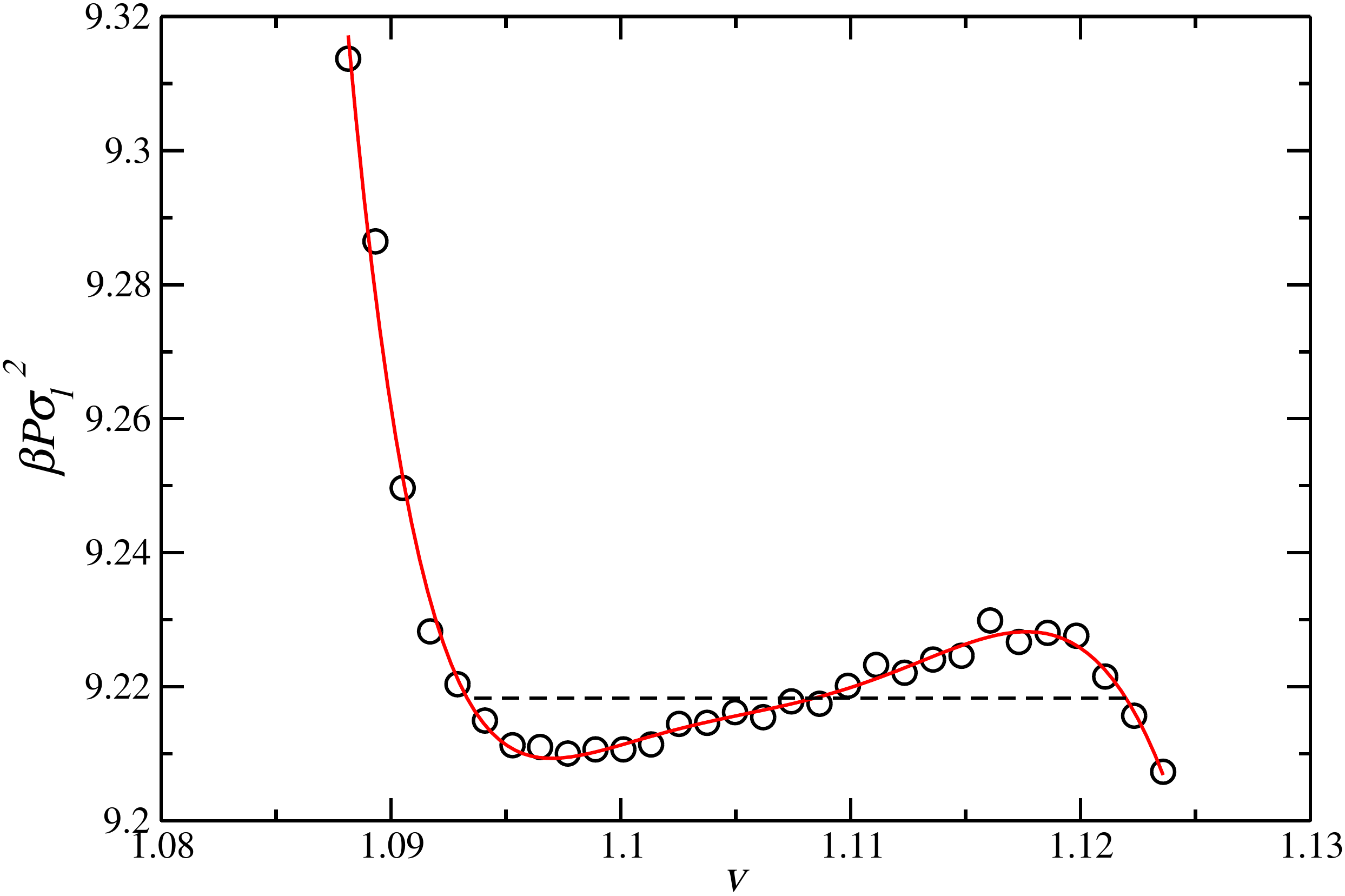}
 \caption{Equation of state for $\xi=2$. Symbols are results from SGCE simulations, the continuous red line is a high-order polynomial fit, while the dashed line is the Maxwell construction.}
 \label{fig:eos_2}
\end{figure}

\begin{figure}[!ht]
 \centering
 \includegraphics[width=10cm]{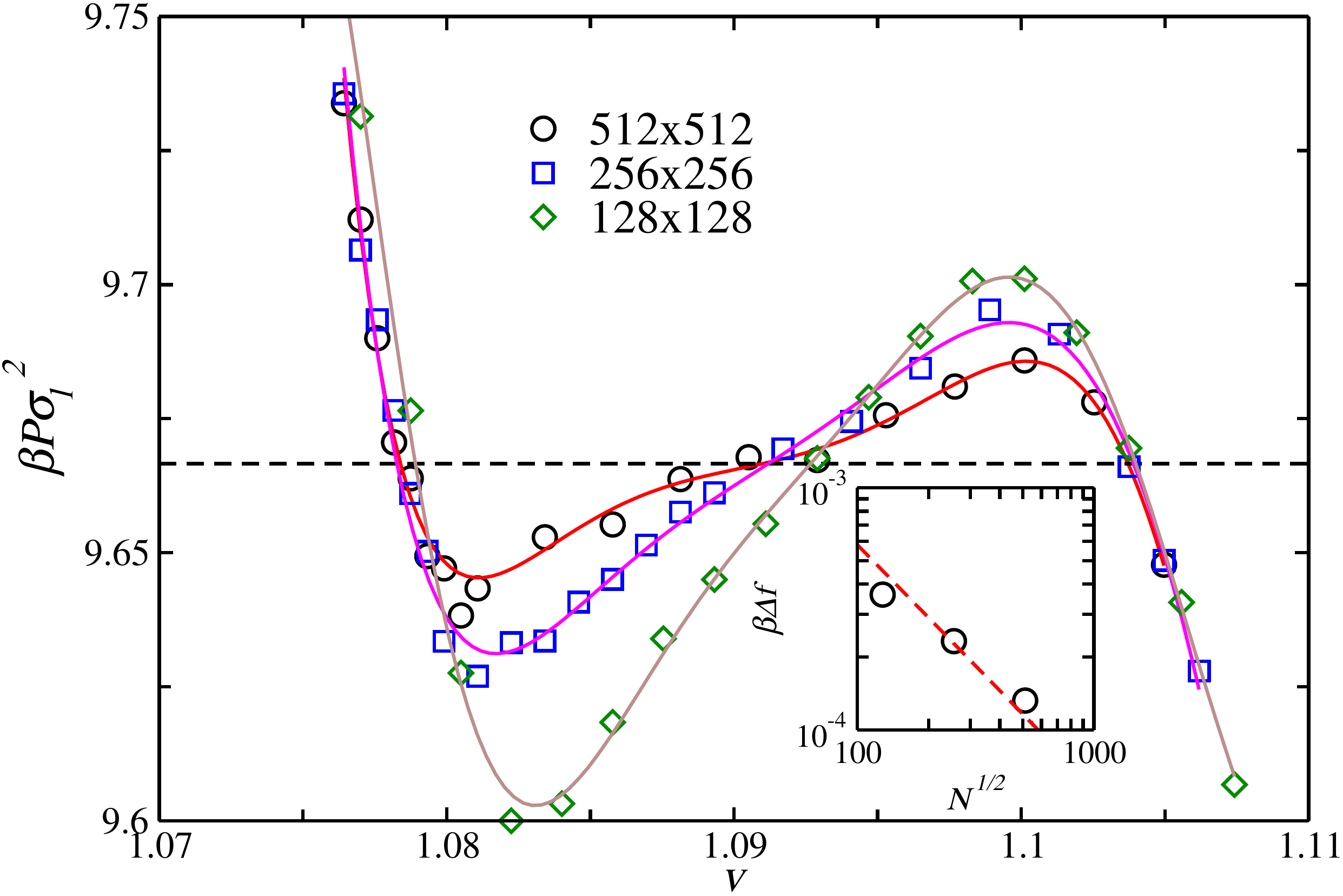}
 \caption{Equation of state for $\xi=22$ for three system sizes. Continuous lines are high-order polynomial fits, while the dashed line is the Maxwell construction for the largest system. The inset shows that the scaling of the interface energy for each disk $\beta\Delta f$, with the dashed line representing $\beta\Delta f\propto N^{1/2}$.}
 \label{fig:eos_22}
\end{figure}

In mean field theories of first-order phase transitions, a so-called van der Waals (vdW) loop refer to the parts of the
equation of state (EOS), $P(\rho)$, that are mechanically unstable. Beyond mean field theory the situation is more
subtle, particularly for simulation. There a phenomenon akin to the vdW loop occurs (and which therefore retains the
name), but it has a very different origin. Specifically the vdW loop observed in simulations arises from interfacial
effects~\cite{binder2012beyond}, theoretical analysis of which predicts that in the limit of large $N$ the area under the loop scales as
the interfacial free energy per disk, $\delta f\propto N^{-1/2}$. An example of such a vdW loop for our system is given
in Fig.~\ref{fig:eos_2}, showing the pressure ($\beta P\sigma_l^2$) as a function of the specific volume $v=1/\rho$ for
simulations at $\xi=2$. The symbols are results from simulations, while the continuous red line is a high-order
polynomial fit to the simulation data. The standard Maxwell construction (dashed line) is used to obtain the coexistence
densities, while the area under the Maxwell construction defines the interface free energy cost of forming an interface,
$\beta\Delta f$. We test the expected scaling in Fig.~\ref{fig:eos_22} where we compare the vdW loop for $\xi=22$ at
three different system sizes, $N=512^2$ (circles), $N=256^2$ (squares), and $N=128^2$ (diamonds). As shown in the inset
of Fig.~\ref{fig:eos_22}, the scaling is indeed compatible with $\beta\Delta f\propto N^{1/2}$ despite the relative
softness of the interface for first-order coexistence in hard disks implying that it is difficult to reach the scaling
limit in this system.

\begin{figure}[!ht]
 \centering
 \includegraphics[width=10cm]{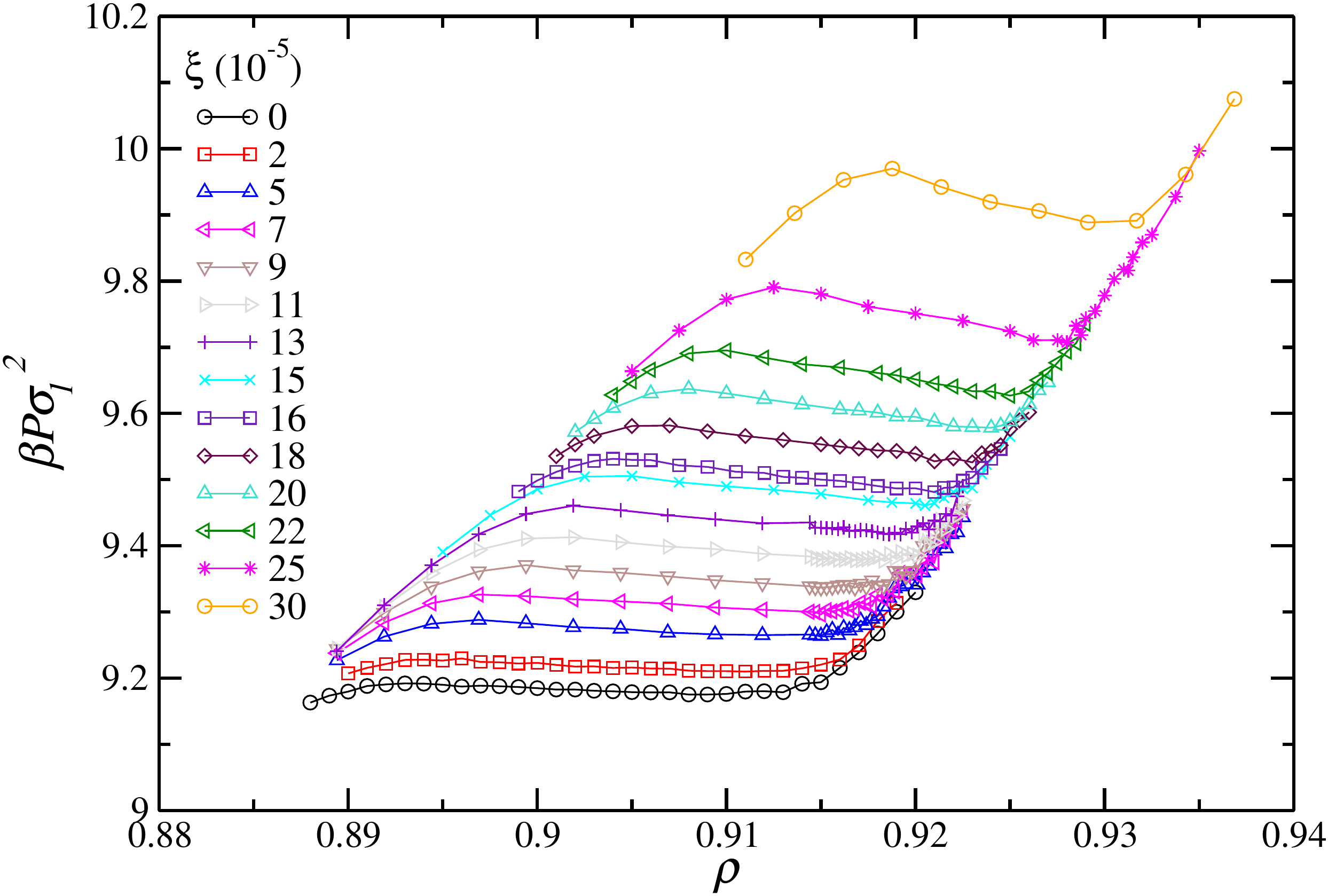}
 \caption{Equation of state for different $\xi$.}
 \label{fig:eos_all}
\end{figure}

In Fig.~\ref{fig:eos_all} we compare the EOS for a large range of $\xi$. With increasing $\xi$ one observes two features: firstly  that the coexistence region shifts to higher densities $\rho$, and pressures $\beta P\sigma_l^2$, and secondly that the free energy barrier between the two phases (which is proportional to the area under the Maxwell construction) grows steadily. This latter feature reflects the increasing difficulty of nucleating the dense phase from the liquid as the concentration of small disks increases towards the eutectic point; it is at the origin of the glass-forming ability of the mixture near the eutectic point.

\section{Continuous Transition: pair correlation functions}\label{sec:c}

To locate the continuous transition, we scan for each $\xi$ a sequence of densities $\rho$.  The sequence commences at the high density boundary of the first-order transition and increases in steps of $\Delta\rho=0.0005$ until the solid phase is reached. For each density considered, we perform $10$ independent simulation runs, each of which yields a measurement of the pair correlation function $g(\Delta\mathbf r)$. We estimate the density of the continuous transition by finding the crossover in the form of $g(\Delta\mathbf r)$ from exponential (short-range, hexatic) to power-law (quasi-long range, solid) behavior.

\begin{figure}[!ht]
 \centering
 \includegraphics[width=10cm]{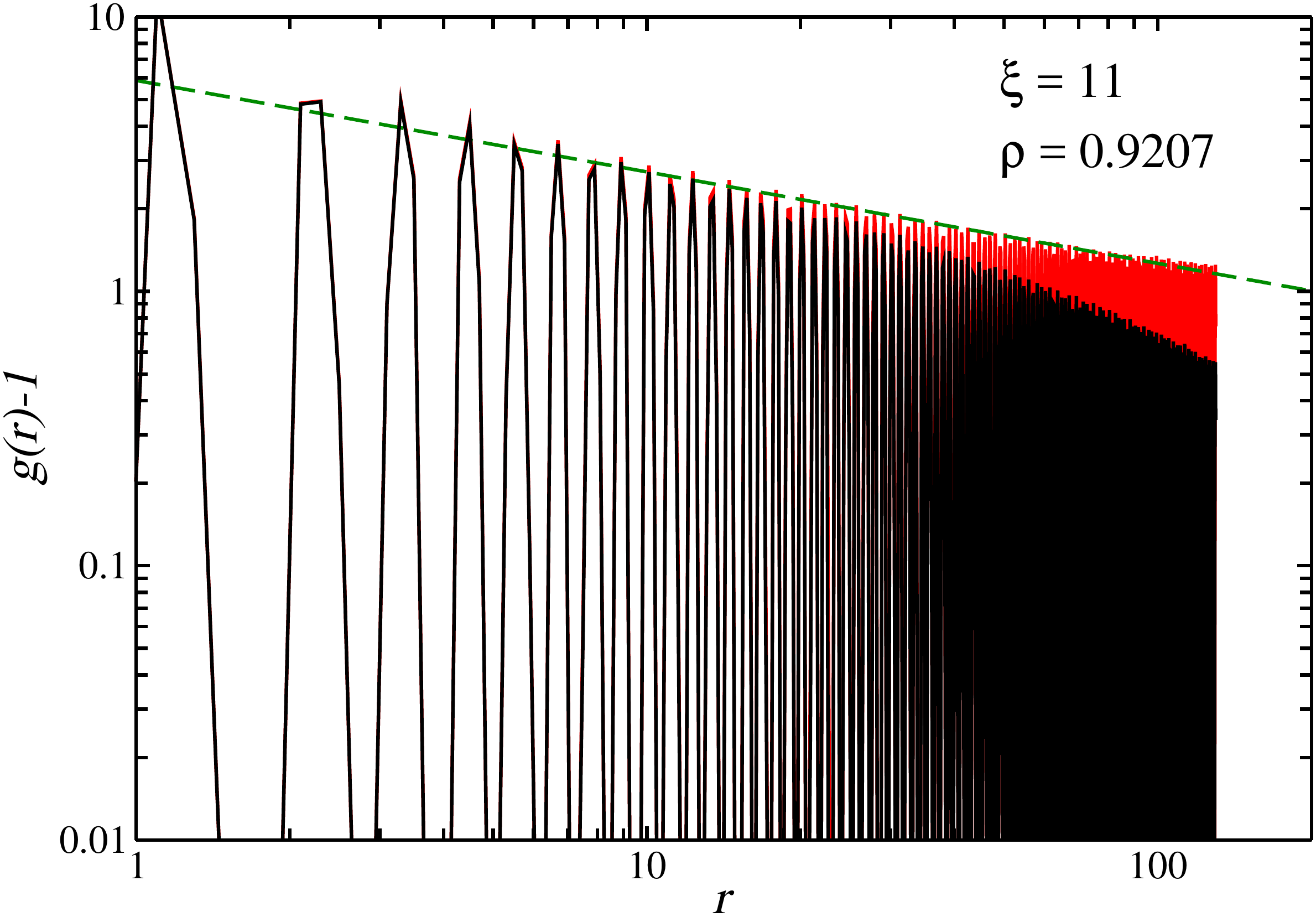}
 \caption{The pair correlation function computed for two independent runs at $\xi=11$ and $\rho=0.9207$: black line for the hexatic phase, red line
for the solid phase, green dashed line is the power-law scaling predicted by KTHNY theory. The system size is $N=256^2$.}
 \label{fig:pair}
\end{figure}

The continuous transition is intrinsically smeared out in a finite-sized system. This is manifest in the fact that even for a state point that is exactly at the transition density, separate independent simulation runs may yield differing forms of the decay of $g(\Delta\mathbf r)$. An example is shown in Fig.~\ref{fig:pair} which displays two independent measurements of the pair correlation function for $\xi=11$ and $\rho=0.9207$, obtained by performing a one dimensional projection of the full two-dimensional pair correlation, $g(\Delta\mathbf r)$, onto the direction of a lattice vector. The black curve shows the result of a run which yielded exponential decay, albeit with a very long correlation length, while the red curve shows the result of a run which yielded power-law decay in the accessible window. The power-law exponent is compatible with $-1/3$ (green dashed line in the figure), which corresponds to the stability limit of the solid phase in KTHNY theory.

Given this phenomenology, we estimate the location of the continuous transition for a given $\xi$ by determining that density for which half of the associated $10$ independent trajectories exhibit a pair correlation function $g(r)$ with a power-law decay (while the rest give exponential decay). The error bars on the transition density in Fig.~1(c) in the main text are determined as follows: the lower-bound indicates the state point at which less than $20\%$ of the $10$ independent runs yield power-law decay of $g(r)$, while the upper bound indicates the state point at which at least $80\%$ of the trajectories have power-law decay. 

\begin{figure}[!ht]
 \centering
 \includegraphics[width=10cm]{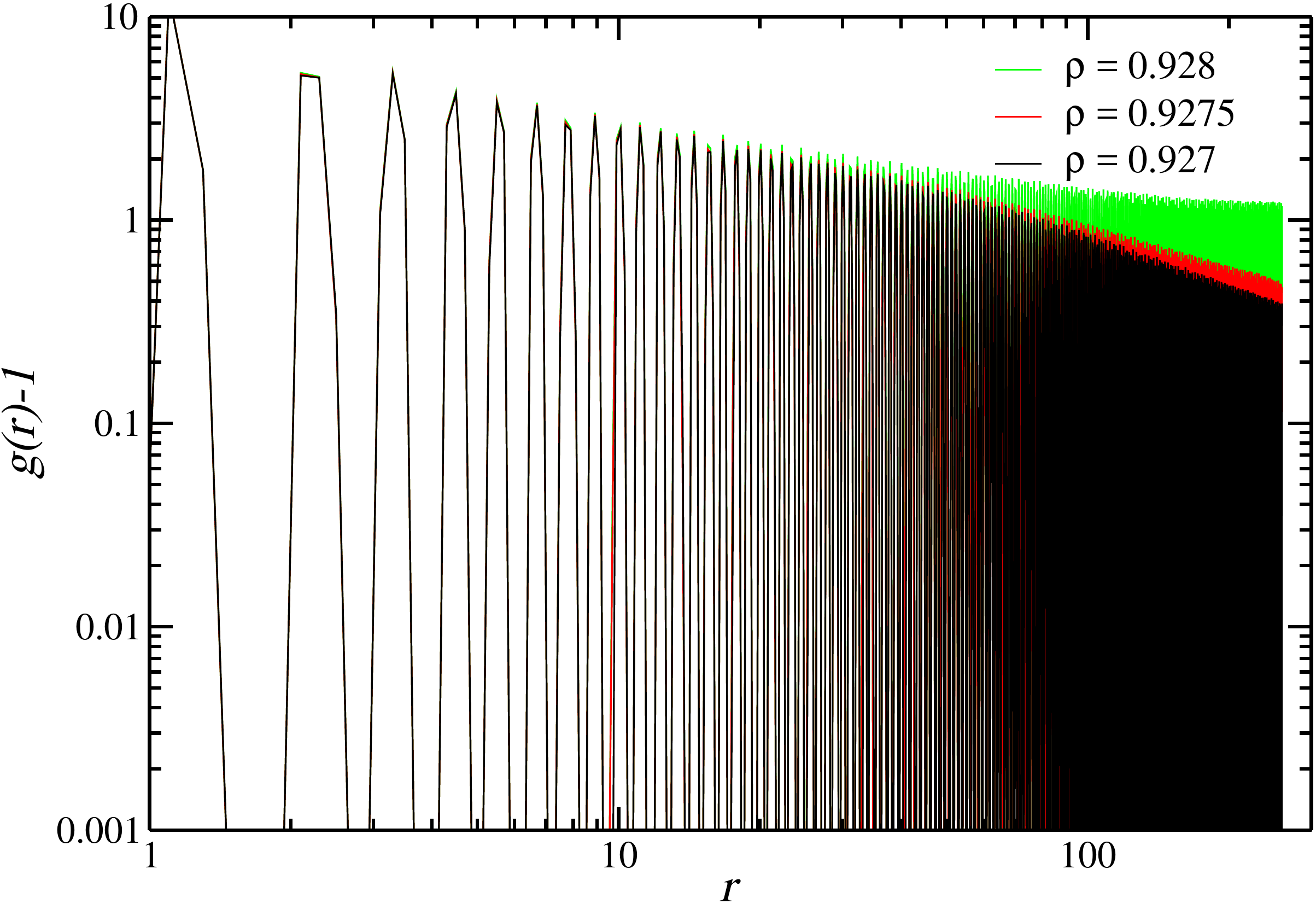}
 \caption{The pair correlation function for three densities covering the continuous transition at $\xi=22$. The system size is $N=512^2$.}
 \label{fig:gr}
\end{figure}

Fig.~\ref{fig:gr} illustrates how the pair correlation function (as measured for our largest system size, $N=512^2$) varies as one crosses the transition density at fixed $\xi=22$.  Data are shown for two densities that straddle the continuous transition, and an intermediate density at the transition point.

\section{Fractionation}\label{sec:fractionation}

\begin{figure}[!ht]
 \centering
 \includegraphics[width=10cm]{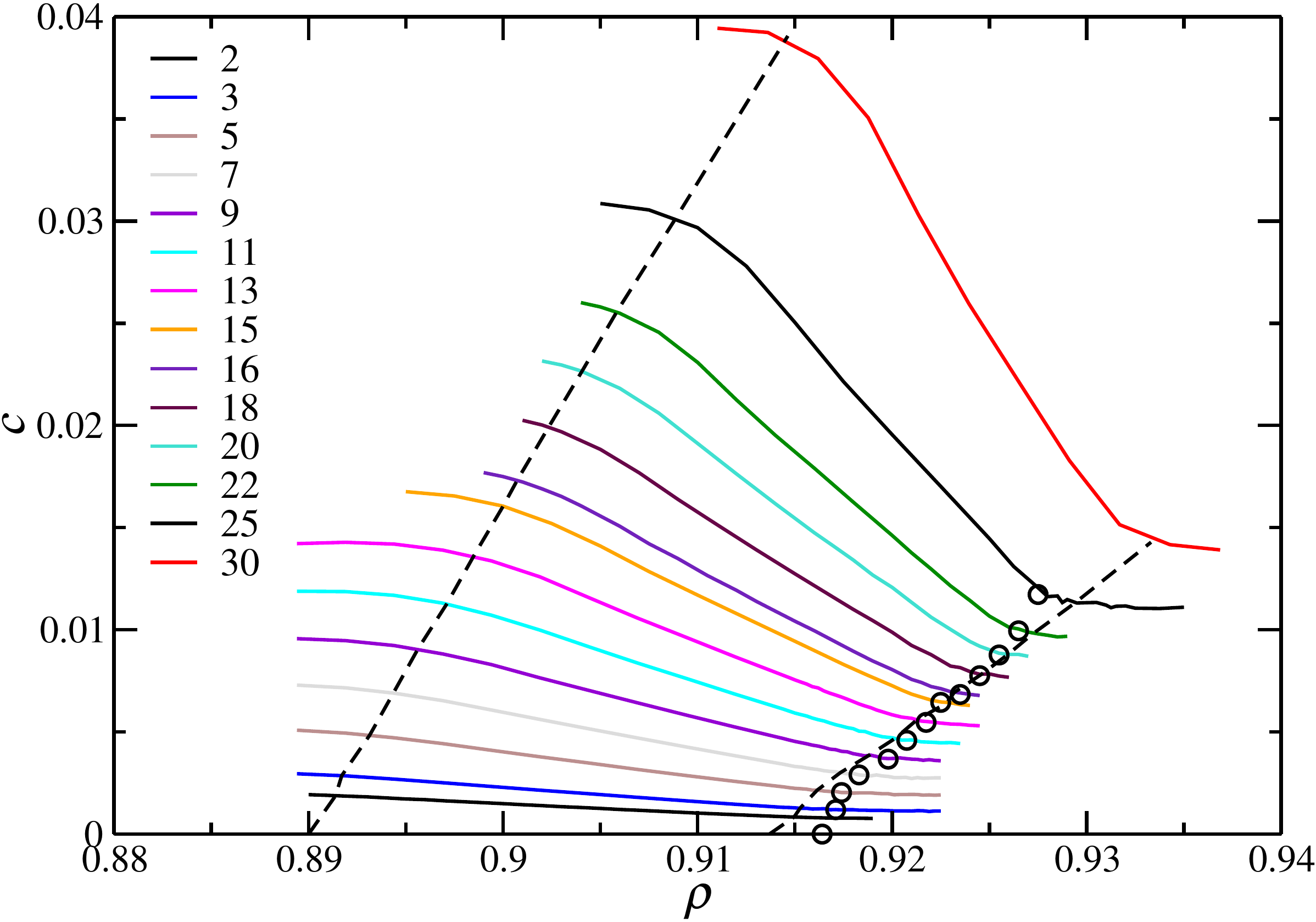}
 \caption{$\rho$ dependence of the concentration $c$ of small disks, for different $\xi$ (continuous lines). Dashed lines represent the coexistence lines of the first-order transition, while open symbols indicate the location of the continuous transition.}
 \label{fig:c}
\end{figure}

Fig.~\ref{fig:c} provides further information on the fractionation of small disks between the different phases. It plots the $\rho$ dependence of the concentration $c$ for different fixed values of $\xi$. Dashed lines mark the coexistence line of the first-order transition, while open symbols give the location of the continuous transition. The results confirm the large  difference in $c$ between the fluid and the hexatic phases seen in fig 1a of the main text, and shows that there is a much smaller difference between the hexatic and solid phases. 

\begin{figure}[!ht]
 \centering
 \includegraphics[width=16cm]{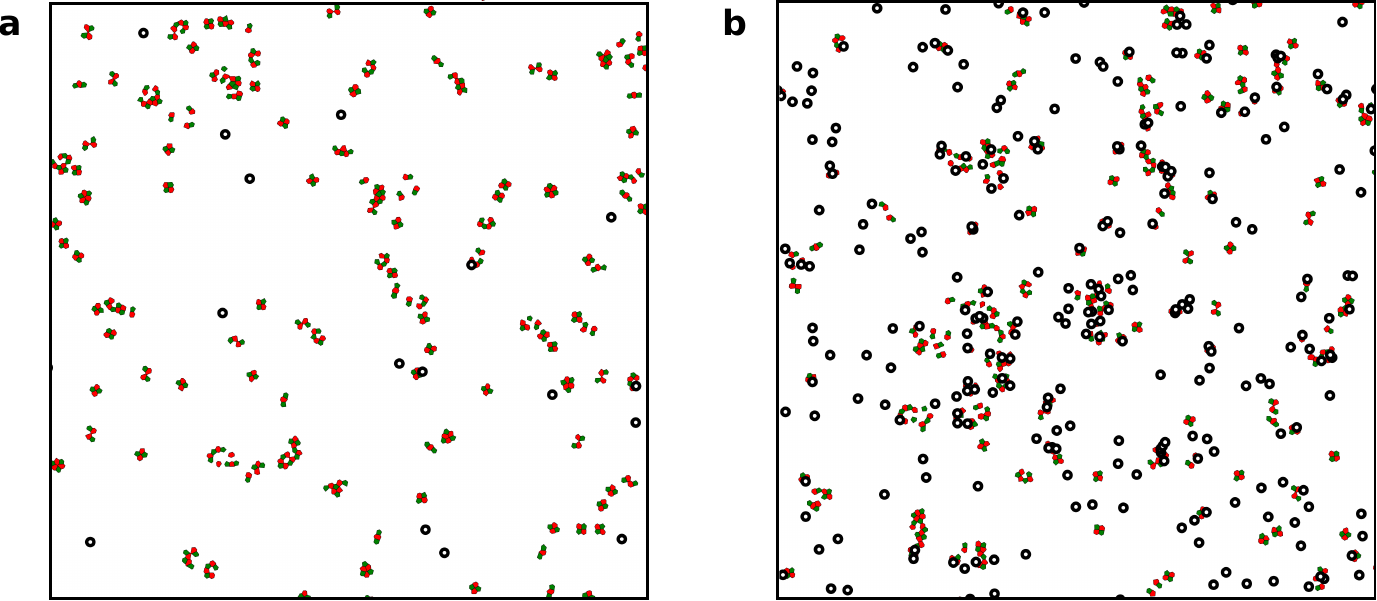}
 \caption{{\bf (a)} Configurational snapshots of a sub-volume of the $N=256^2$ system size for  $\xi=3$, $\rho=0.9171$. Small disks are shown at twice their real size (large disks are omitted).
Also shown are the Voronoi cells representing topological defects with 5 (green cells) and 7 (red cells) neighbors. {\bf (b)} As for (a) but for $\xi=30$, $\rho=0.9343$.}
 \label{fig:voronoi_d}
\end{figure}

It is illuminating to visualize the configurations of small particles, particularly in relation to other structural features such as solid phase defects. The snapshots in Fig.~\ref{fig:voronoi_d}(a) and (b) show examples of the distribution of small disks for $\xi=3$ and $\xi=30$ respectively at densities corresponding to the stability limit of the solid phase. The plots (which for clarity display only a quarter of the full $N=256^2$ particle system) confirm that $c$ is much
larger for $\xi=30$ than for $\xi=3$. They also reveal (as noted in the main text) that the small particles are much more clustered for $\xi=30$ and furthermore there is a high correlation between these small particle clusters and the location of clustered topological
defects, i.e. grain boundaries.

\begin{figure}[!ht]
 \centering
 \includegraphics[width=10cm]{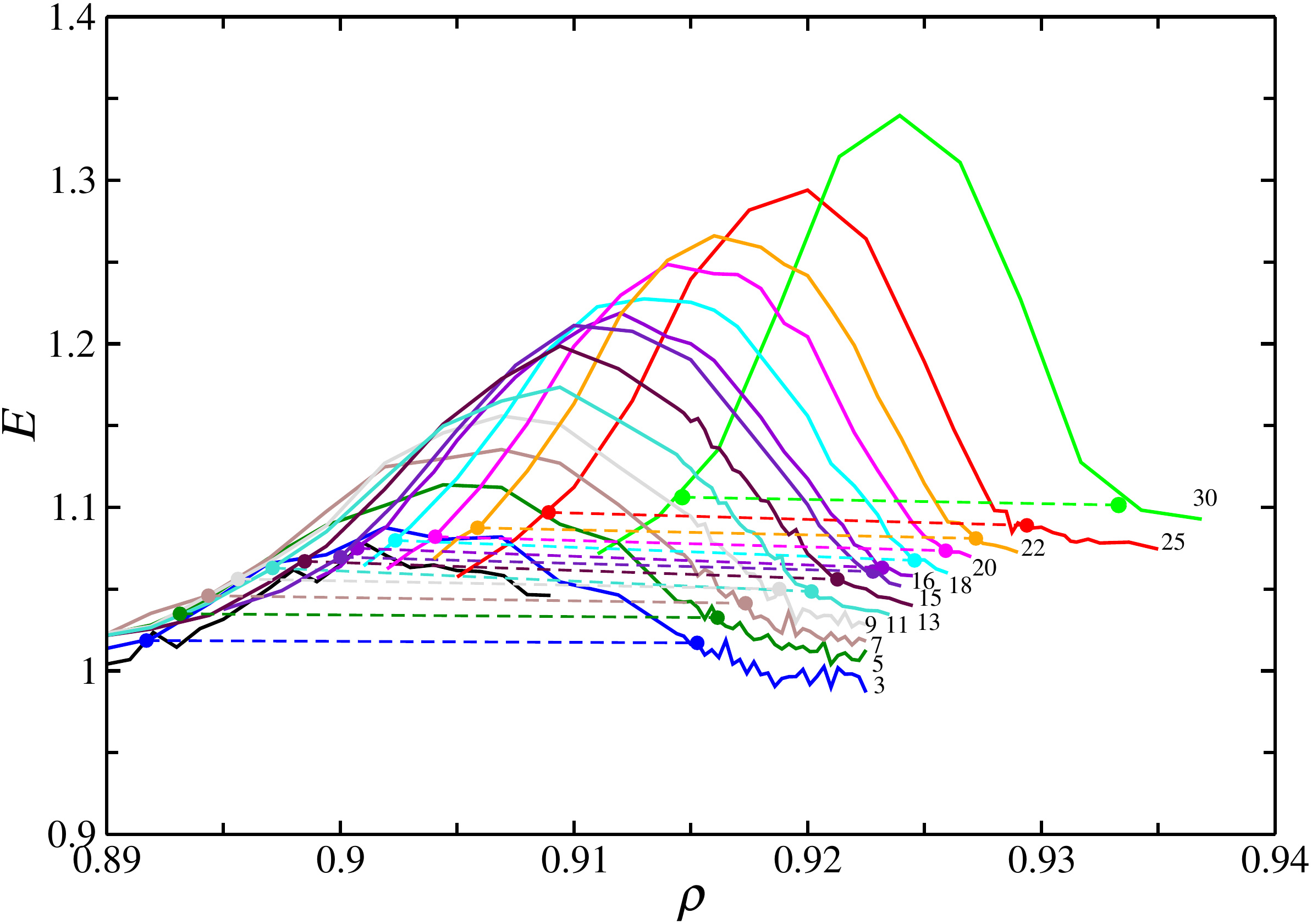}
 \caption{Density dependence of the Pielou evenness index for various fugacities ranging from $\xi=2$ (bottom curve) to $\xi=30$ top curve. The full symbols on each curve mark the coexisting densities of the first-order transformation, while the dashed line represents the coexistence region.}
 \label{fig:pielou}
\end{figure}

As explained in the main text, the Pielou index ($E$) quantifies the degree of clustering of the small disks. In Fig.~\ref{fig:pielou} we plot this index for a wide range of values of $\xi$. Each curve corresponds to a different $\xi$, and the corresponding coexistence densities are marked with full symbols on the curve, and by a dashed line connecting these symbols. Following one curved from low to high density, the $E$ index increases rapidly on entering the coexistence region, before attaining a maximum having $E>1$ in the middle of this region. This peak reflects the combined effects of fractionation and phase separation: the preference of the small disks for the liquid phase artificially enhances the $E$-index when the liquid phase has a fractional volume $<1$ i.e. when it occupies a sub-volume of the system. Beyond the middle of the coexistence region, the artificial enhancement of $E$ diminishes in tandem with the fractional volume of the liquid phase. On reaching the high density boundary of the coexistence region, $E$ continues to decreases, until on entering the solid region its $\rho$-dependence flattens out. For each $\xi$, the $E$ 
index of the two coexisting densities (full symbols) is almost the same, as discussed in the main text. This implies that the first-order transition involves two phases that, despite their very different $c$, have the same evenness, as discussed in the main text. 

\section{Computation of the elastic constants}\label{sec:elastic}

Here we outline the procedure for the calculation of the elastic constants. For a more in-depth discussion we refer the reader to Refs.~\cite{sengupta2000elastic,sengupta2000elastic2}.

The first step involves the calculation of the elastic constants of a defect-free triangular solid of $N=3120$ disks in a quasi-square box. The most probable topological defect is the formation of a dislocation pair of the smallest Burger's vector, where a configuration of four disks having $7$-$5$-$7$-$5$ neighbors appears. This can be avoided by rejecting all Monte Carlo moves which result in a disk having a number of neighbors different from $6$. Neighbors are defined by constructing the Delaunay triangulation which is dual to the Voronoi tessellation of the plane. In this restricted ensemble, the linearized stress tensor can be computed
$$
\epsilon_{ij}=\frac{1}{2}\left(\frac{\partial u_i}{\partial R_j}+\frac{\partial u_j}{\partial R_i}\right)\:,
$$
where $i$,$j$ refer to the components $x$ and $y$ of a Cartesian coordinate system in the body reference frame (Lagrangian representation of the stress tensor), $u_i$ is the displacement vector between the position of a disk and its reference lattice position, and $R_i$ is the lattice vector corresponding to that disk.
Fluctuation in the stress tensor define the compliance matrix
$$
S_{ijkl}=\left\langle\epsilon_{ij}\epsilon_{kl}\right\rangle \:.
$$
For the calculation of the elastic constants the relevant components of the compliance matrix are $S_{11}=\left\langle\epsilon_{xx}\epsilon_{xx}\right\rangle$ and $S_{12}=\left\langle\epsilon_{xx}\epsilon_{yy}\right\rangle$. From these, one can build the linear combinations $S_{++}=2(S_{11}+S_{12})$ and $S_{--}=2(S_{11}-S_{12})$, and obtain the bulk ($B$) and effective shear modulus ($\mu$) as
\begin{eqnarray}
 \beta B &=& \frac{1}{2S_{++}} \nonumber\:, \\
 \beta \mu &=& \frac{1}{2S_{--}} \nonumber\:.
\end{eqnarray}

The system size independent values of $S_{++}$ and $S_{--}$ are obtained via finite-size scaling. For each simulation of
size $L$, we measure the fluctuations of the strain tensor in sub-boxes of size $L_b$. The $L_b/L$ dependence of these
fluctuations can be computed via a quadratic Landau functional. Fitting simulation results with the appropriate
functional form~\cite{sengupta2000elastic} allows the determination of the size independent fluctuations $S_{++}$ and
$S_{--}$. In our case, the fluctuations at system size $L_b$ are obtained by averaging over $100$ sub-boxes randomly
placed within the system, and over $10$ independent runs. The fitting procedure is then applied for system sizes
$L_b/L<0.5$, which gave the best agreement with the values of the bulk modulus independently computed via the equation
of state (see below).

To compute the elastic constants of mixtures, the calculations above are run in the SGCE ensemble, where small disks
substitutionally occupy a lattice position of the solid, and swap moves allow for fast sampling of independent
configurations. To check for ensemble-dependence, we have run the same calculations also in the NVT ensemble, with the
concentration of small disks fixed at the equilibrium values, and observed no difference in the results.

\begin{figure}
 \centering
 \includegraphics[width=10cm]{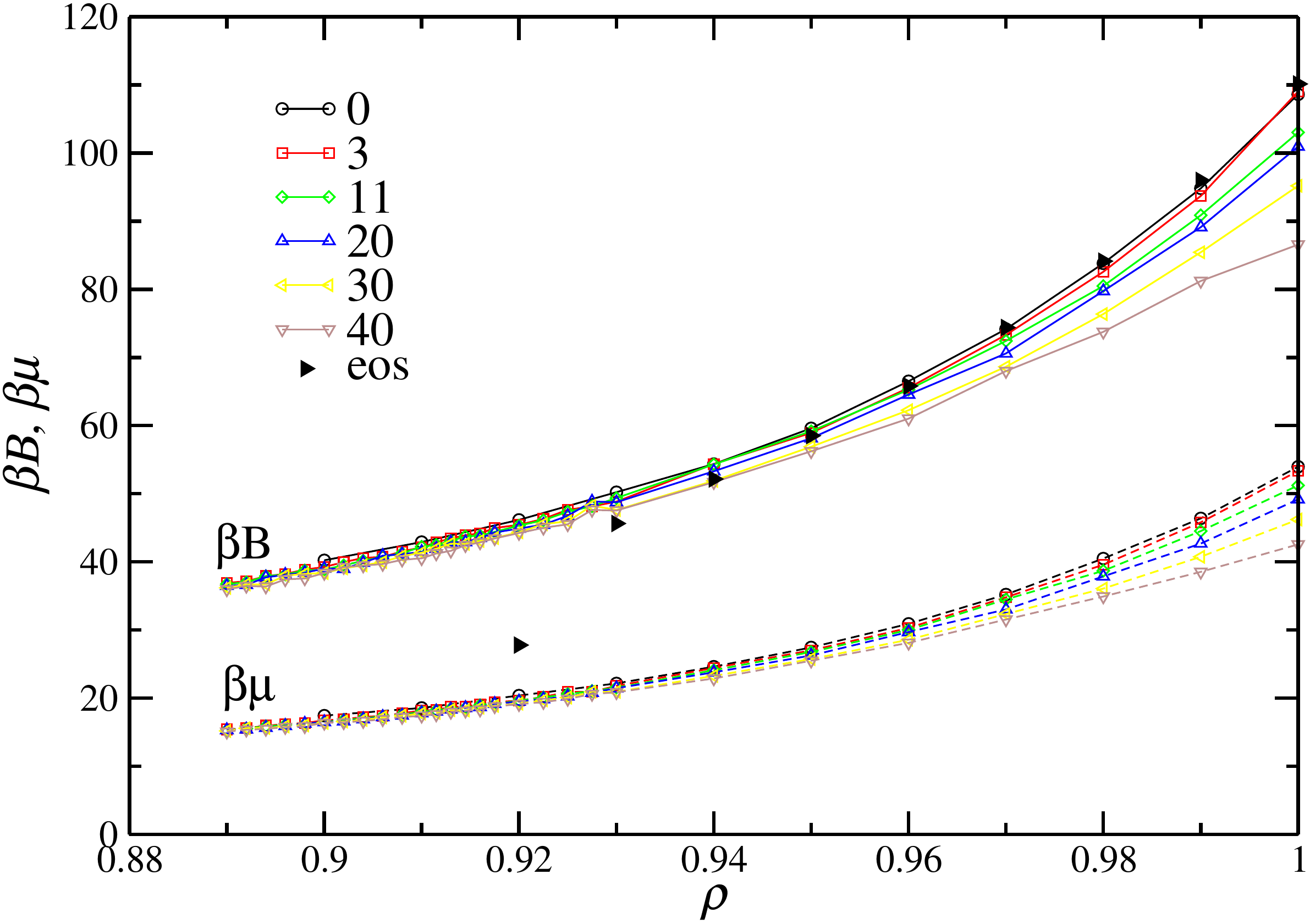}
 \caption{Bulk ($B$) and effective shear ($\mu$) modulus as a function of density $\rho$, for various values of the fugacity fraction $\xi$ (symbols). The full symbols plot the bulk modulus computed from the equation of state of the unconstrained system.}
 \label{fig:elastic}
\end{figure}

Fig.~\ref{fig:elastic} plots the bulk ($B$) and effective shear ($\mu$) modulus for the monodisperse case ($\xi=0$) and for mixtures with fugacity fraction ($\xi$). As a consistency check we can compare the results of the bulk modulus for the monodisperse system (open black circles), with an independent estimate of the bulk modulus via the following relation
$$
B=\rho\frac{\partial P}{\partial\rho}\:.
$$
The corresponding bulk modulus, computed with independent unconstrained simulations, is plotted in Fig.~\ref{fig:elastic} with full triangle symbols. We observe that, at high $\rho$, there is excellent agreement between the bulk modulus computed via the strain tensor (open circles) and the one computed from the equation of state (full symbols). This agreement breaks down at low $\rho$, because of crystalline defects that proliferate in the unconstrained simulations, and eventually cause the crystal to melt.

With the values of $B$ and $\mu$ of Fig.~\ref{fig:elastic} we can obtain the Lame' coefficient $\lambda=B-\mu$, and from this compute the Young elastic modulus K
$$
K=\frac{8}{\sqrt{3}\rho}\frac{\mu}{1+\mu/(\lambda+\mu)}\:.
$$

In order to apply the KTHNY recursion relations, we need an estimate of the fugacity of dislocation pairs, $y=\exp(-E_c)$, defined in terms of the core energy of the dislocation, $E_c$. We can obtain this from the dislocation pair probability ($p_d$) via the following relation
\begin{equation}\label{eqn:pd}
 p_d=\exp(-2E_c)\frac{2\pi\sqrt{3}}{K/8\pi-1}I_0\left(\frac{K}{8\pi}\right)\exp{\left(\frac{K}{8\pi}\right)}\:,
\end{equation}
where $I_0$ is a Bessel function.

\begin{figure}
 \centering
 \includegraphics[width=10cm]{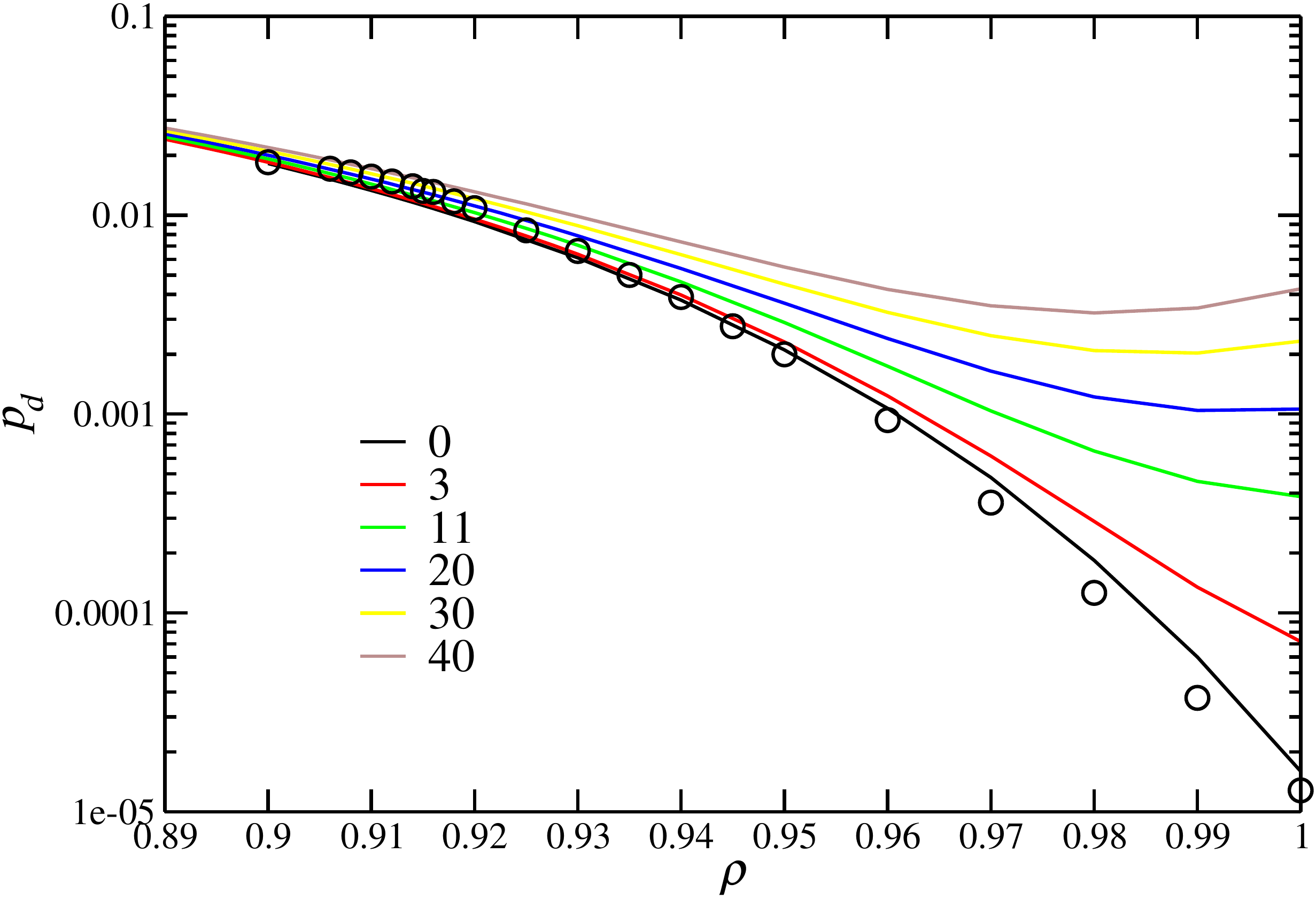}
 \caption{Probability of dislocation pairs $p_d$ for different values of $\xi$ (continuous lines). The open symbols plot the site-probability $N_i/N$ of finding a lattice site in a dislocation pair.}
 \label{fig:pd}
\end{figure}

Following Ref.~\cite{sengupta2000elastic2}, we can assume that $p_d$ is proportional to the acceptance probability of the constrained MC
simulations that conserve the topology of the Delaunay triangulation. To fix the proportionality constant, for the
monodisperse case, we compute the site probability $N_i/N$ of finding a dislocation pair in unconstrained simulations,
and compare it to the acceptance probability of constrained MC simulations. This is shown in Fig.~\ref{fig:pd} where we
plot the $\rho$ dependence of the site probability $N_i/N$ for the monodisperse case (open black symbols), and the
acceptance probability of the constrained MC simulations after a simple rescaling (continuous black line). We can see
that the two quantities have a very similar density dependence. We then compute $p_d$ from the acceptance rate of
constrained simulations of binary mixtures, and observe that $p_d$ increases with increasing $\xi$ (continuous lines in
Fig.~\ref{fig:pd}). At high fugacity fraction ($\xi\gtrsim 20$) we observe that $p_d$ is a non-monotonous function of
$\rho$, and starts increasing at high values of $\rho$. This is due to the re-entrant melting of the solid phase at high
density, which is typical of eutectic systems below the eutectic point.

From the values of $p_d$ in Fig.~\ref{fig:pd}, we use Eq.~\ref{eqn:pd} above to extract $E_c$, and finally $y$.

The KTHNY recursion relations give the renormalized value of the Young modulus ($K$) and the fugacity of dislocations ($y$) in the limit $l\rightarrow\infty$
\begin{eqnarray}
 \frac{\partial K^{-1}}{\partial l} & = & 3\pi y^2\exp{(K/8\pi)}\left[\frac{1}{2}I_0\left(\frac{K}{8\pi}\right)-\frac{1}{4}I_1\left(\frac{K}{8\pi}\right)\right] \:,\nonumber \\
 \frac{\partial y}{\partial l} & = & \left(2-\frac{K}{8\pi}\right)y+2\pi y^2\exp{(K/16\pi)}I_0\left(\frac{K}{8\pi}\right) \:,\nonumber
\end{eqnarray}
which are solved with a Runge-Kutta scheme with $\Delta l=0.001$ until $l=100$.

\begin{figure}
 \centering
 \includegraphics[width=10cm]{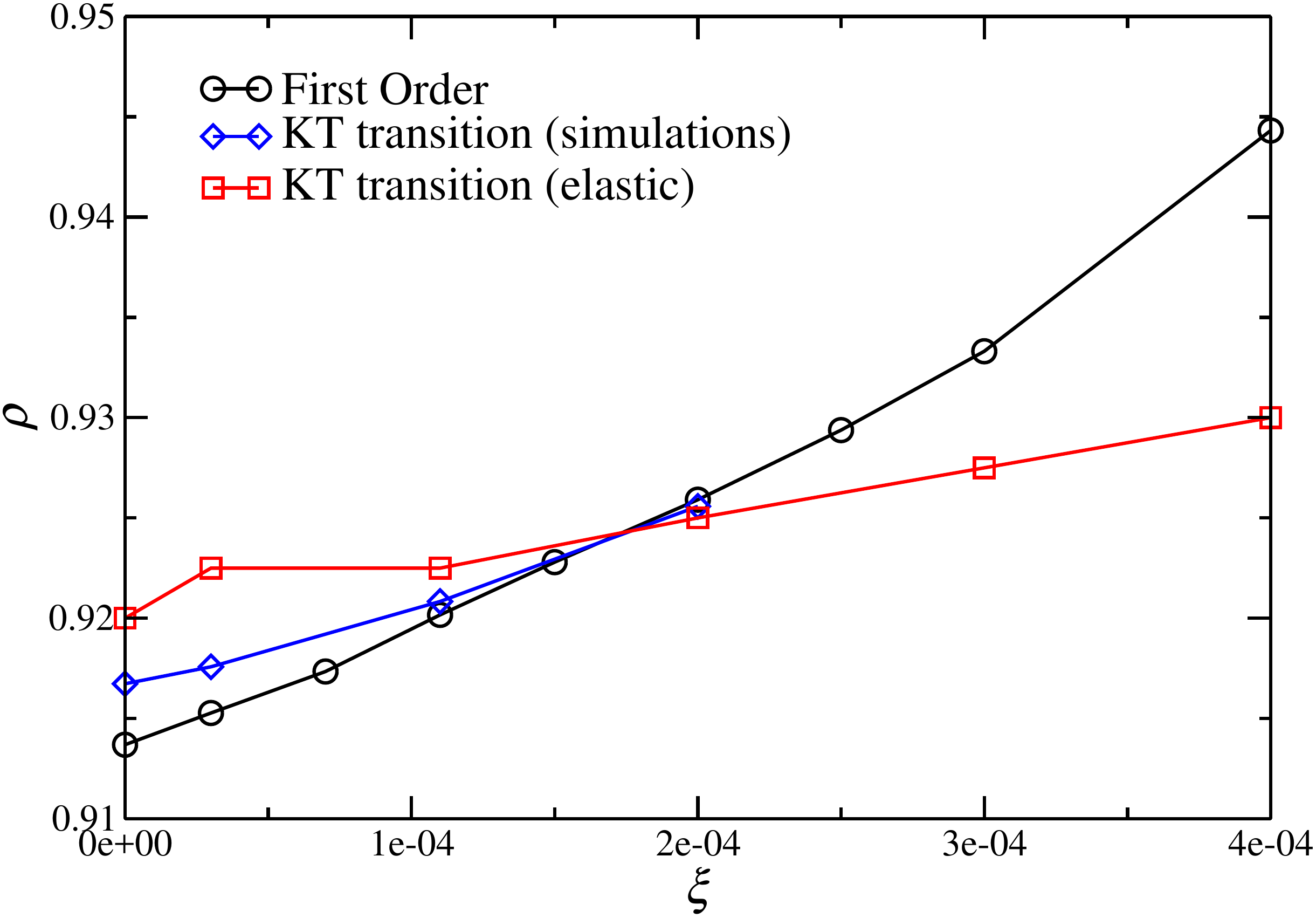}
 \caption{Transitions points as a function of $\xi$. Circle symbols represent the $\rho$ of the first-order transition points (the high-$\rho$ branch); diamonds symbols mark the location of the continuous transition estimated with MC simulations; squares symbols mark the continuous transition estimated with KTHNY theory after calculation of the elastic constants.}
 \label{fig:transitions}
\end{figure}
At the transition, the values of $K$ jump discontinuously to zero. The transition points computed through KTHNY theory are plotted as square symbols in Fig.~\ref{fig:transitions}. The figure also plots the location of the first-order (circle symbols) and continuous transitions (diamond symbols) obtained with MC simulations. Both calculations predict the shrinking of the hexatic phase stability window, and a melting transition that is first-order at high $\xi$.

\end{document}